\documentstyle[prc,preprint,aps]{revtex}
\begin{document}
\draft
\tightenlines
\title{Neutrons from multiplicity-selected La-La and Nb-Nb \\
collisions at 400A MeV and La-La collisions at 250A MeV} 
\author {M.\ M.\ Htun\cite{htun}$^{1}$, R.\ Madey$^{1}$, W.\ M.\ Zhang$^{1}$, 
M.\ Elaasar\cite{mae}$^{1}$, D.\ Keane$^{1}$, B.\ D.\ Anderson$^{1}$, A.\ R.\ 
Baldwin$^{1}$,  
J.\ Jiang\cite{jji}$^{1}$, A. Scott$^{1}$, Y.\ Shao\cite{yip}$^{1}$, J.\ W.\ 
Watson$^{1}$, 
K.\ Frankel$^{2}$, L.\ Heilbronn$^{2}$, G.\ Krebs$^{2}$, M. A.\ McMahan$^{2}$, 
W. Rathbun$^{2}$, J.\ Schambach\cite{jos}$^{2}$, G.\ D.\ Westfall$^{3}$, S.\ 
Yennello\cite{yen}$^{3}$, C.\ Gale$^{4}$, and J.\ Zhang$^{4}$}
\address{$^{1}$Kent State University, Kent, Ohio 44242 \\ 
$^{2}$Lawrence Berkeley Laboratory, Berkeley, California 94720 \\ 
$^{3}$Michigan State University, East Lansing, Michigan 44828 \\
$^{4}$McGill University, Montr{\'e}al, Qu{\'e}bec, Canada H3A 2T8 }
\maketitle
\vspace{2in}
\newpage
\begin{abstract}

Triple-differential cross sections for neutrons from high-multiplicity La-La 
collisions at 250 and 400 MeV per nucleon and Nb-Nb collisions at 400 MeV 
per nucleon were measured at several polar angles as a function of the 
azimuthal angle with respect to the reaction plane of the collision.  The 
reaction plane was determined by a transverse-velocity method with the 
capability of identifying charged-particles with $Z=1$, $Z=2$, and $Z > 2$.  
The flow of neutrons was extracted from the slope at mid-rapidity of the
curve of the average in-plane momentum vs the center-of-mass rapidity.
The {\em squeeze-out} of the participant neutrons was observed 
in a direction normal to the reaction plane in the normalized momentum 
coordinates in the center-of-mass system.  Experimental results of the neutron 
squeeze-out were compared with BUU calculations.  The polar-angle dependence 
of the maximum azimuthal anisotropy ratio $r(\theta)$ was found to be 
insensitive to the mass of the colliding nuclei and the beam energy.  
Comparison of the observed polar-angle dependence of the maximum azimuthal 
anisotropy ratio $r(\theta)$ with BUU calculations for {\em free} neutrons 
revealed that $r(\theta)$ is insensitive also to the incompressibility modulus 
in the nuclear equation of state.  

\end{abstract}
\vspace{1in}
\narrowtext
\section{Introduction}

A primary goal of relativistic heavy-ion physics is to extract the 
equation-of-state (EOS) of nuclear matter.  The EOS is the relationship 
between density, temperature, and pressure for nuclear matter 
\cite{ck86,cs86,sg86,sto86}.  Various experiments 
\cite{bea85,bea86,do85,dos85,dos86,gus84,gut89,gut90,rit85,lei93,dos87} have 
been undertaken in an effort to extract the EOS.  In central collisions, the 
overlap regions of the target and the projectile form a compression zone of 
high density.  Particles in this zone experience a deflection or outward flow 
from the interaction region.  Measurements \cite{dos86,gus84,gut89,dos87} with 
the Plastic Ball-Wall detector \cite{bab82} revealed three collective flow 
effects:~~the {\em bounce-off} of the spectators, the {\em side-splash} of the 
participants, and the {\em squeeze-out} of the participants in directions 
normal to the reaction plane.  Measurements of the collective momentum flow in 
the collisions, including the flow angle \cite{gus84,rit85} and the average 
transverse momentum in the reaction plane \cite{bea86,do85,dos86,dos87,kea88}, 
were studied with a view toward extracting the nuclear EOS.  

Zhang {\em et 
al.} \cite{au93} investigated the collective flow of neutrons correlated to 
the azimuthal distributions for semicentral Au-Au collisions at beam energies 
of 150, 250, 400, and 650 MeV per nucleon.   Elaasar {\em et al}. 
\cite{nb93,ela94} reported the first azimuthal distributions of 
triple-differential cross sections for neutrons emitted in Nb-Nb collisions
at 400A MeV and examined the maximum azimuthal anisotropy ratios of 
these neutron cross sections as a probe of the nuclear equation 
of state by comparison with Boltzmann-Uehling-Uhlenbeck (BUU)
calculations for {\it free} neutrons with a momentum-dependent interaction. 
Madey {\it et al.}\cite{mad93} extended this comparison to 400A MeV Au-Au
collisions and found that the maximum azimuthal anisotropy ratio does not depend
on the mass of the colliding nuclei.
Welke {\em et al.} \cite{wel88} defined the 
maximum azimuthal anisotropy ratio $r(\theta)$ as a ratio of the maximum 
triple-differential cross section to the minimum triple-differential cross 
section at each polar angle.  

In this paper, we report measurements of triple-differential cross sections of 
neutrons emitted in high-multiplicity La-La collisions at 250 and 400A MeV 
(and Nb-Nb collisions at 400A MeV) as a function of the azimuthal angle with 
respect to the reaction plane for several polar angles.  From these data  
and our prior data \cite{au93} for Au-Au collisions at 400A MeV, 
we examined the sensitivity of the {\em maximum 
azimuthal anisotropy ratio} $r(\theta)$ to the mass of three 
systems [viz., Nb-Nb, La-La, and Au-Au] and to the beam energy
from 250 to 400A MeV for the La-La system;   
we extracted the flow  
\cite{dos86} of neutrons from the slope at mid-rapidity of the curve of the 
average in-plane momentum versus the neutron center-of-mass rapidity; and
we observed the {\it out-of-plane squeeze-out} of neutrons in 
three systems [viz.,
250 and 400A MeV La-La, and 400A MeV Nb-Nb].

\section{Experimental Apparatus}

Based on experience gained from experiment 848H in 1988 \cite{au93,nb93}, an 
extension of experiment 848H was performed with the Bevalac at the Lawrence 
Berkeley Laboratory (LBL) with the improved experimental arrangement shown in 
Fig.~\ref{fig:expt}.  
This experiment was improved over the original version by 
adding the capability of identifying charged-particles with $Z=1$, $Z=2$, and 
$Z > 2$. With this improvement, data were obtained for La-La collisions at
250 and 400A MeV and for Nb-Nb collisions at 400A MeV.

A beam telescope, consisting of two scintillators S$_{1}$ and S$_{2}$, was 
used to select and count valid projectile beam ions and to provide a fiducial 
timing signal for the measurement of the flight times to the time-of-flight 
(TOF) wall detectors and to the neutron detectors.  Scintillator S1 was 
positioned 13.04~m upstream of the target in the beam.  Scintillator 
S2 was placed immediately before the target.  A total of 16 mean-timed 
neutron detectors \cite{bm80,cam79,mad78,mad83} were placed at angles from 
3$^{\circ}$ to 80$^{\circ}$ with flight paths ranging from 6 to 8.4 m.  
Table~\ref{tab:neuss} shows the width, polar angle, and flight path of each 
neutron detector used in this experiment.  The width of the neutron detector 
and the flight path at each angle were selected to provide approximately equal 
counting rates and energy resolutions for the highest energy neutrons at each 
angle.  To avoid detecting charged particles in neutron detectors, 9.5-mm 
thick anticoincidence detectors were placed immediately in front of the 
neutron detectors.  The TOF of each detected neutron was determined by 
measuring the time difference between the detection of a neutron in one of the 
neutron detectors and the detection of a projectile ion in the beam telescope.  

The azimuthal angle of the reaction plane was determined from the 
information given by a time-of-flight (TOF) wall, which consisted of 184 
plastic scintillators, each with a thickness of 9.5~mm.  The overall dimensions 
of the TOF wall were 5-m wide and 4.5-m high.  The plastic wall was shaped as 
an arc around the target, and covered angles of $\pm 37^{\circ}$ relative to 
the beam line.  The flight paths of the TOF wall detectors varied from about 
4.0 to 5.0~m.  

To assess the background, steel shadow shields were positioned approximately 
half-way between the TOF wall and the neutron detectors in four different 
configurations (see Table~\ref{tab:neuss}).  The resulting spectra were then 
subtracted from the ones without shadow shields.  We used a method introduced 
by Zhang {\em et al.} \cite{au93} to correct for an over-estimation of 
backgrounds by this shadow-shield subtraction technique.  

\section{Data Analysis}

In order to suppress the background arising from collisions of beam ions 
and air molecules between the target and the TOF wall, an appropriate 
multiplicity-cut had to be made.  Figure~\ref{fig:mult} shows the 
charged-particle multiplicity spectra with (solid line) 
and without (dashed line) the target for La-La 
and Nb-Nb collisions at 400A MeV and La-La collisions at 250A MeV.  A proper 
multiplicity-cut was chosen by comparing the charged-particle multiplicity 
spectra with and without the target in place.  To make meaningful comparisons 
of the data for La-La and Nb-Nb collisions at 400A MeV and La-La collisions at 
250A MeV, the multiplicity cuts chosen should correspond to approximately the 
same range in normalized impact parameter while keeping the background 
contaminations (from collisions of beam ions with the air or material in the 
beam telescope) below 5\%.  For La-La collisions at 400 and 250A MeV, the 
appropriate charged-particle multiplicity cuts were found to be 34 and 30, 
respectively, whereas a multiplicity cut of 25 was determined for 
Nb-Nb collisions at 400A MeV.  
From the relationship between the total number of interactions expected from 
the geometric cross section of the system and with the well-known assumption 
\cite{nb93,cav90} of a monotonic correlation between the impact 
parameter and the fragment multiplicity, the maximum impact parameter 
normalized to the radii of the projectile and target, $b_{max}/(R_{p}+R_{t})$, 
was found to be about 0.5 for the above multiplicity cuts.  The method of 
extracting the maximum-normalized impact parameter $b_{max}/(R_{p}+R_{t})$ can 
be found in reference~\cite{ela94}.  

\subsection{Determination of Reaction Plane} \label{sec:rplane}

To study flow, it is necessary to determine the reaction plane, ({\em 
i.e.},~~the $xz$-plane), where {\^x} is in the direction of the impact 
parameter of the collision, and {\^z} is in beam direction.  The reaction plane 
was determined by a modified version \cite{fzg87} of the transverse momentum 
method \cite{do85}.  The charged particles emitted from the collisions between 
beam ions and the target were detected by the TOF wall detectors.  
The azimuthal angle $\Phi_R$ of the reaction plane was estimated from the vector
$\vec{Q}^\nu$, the weighted sum of the transverse velocity vectors $\vec{V}^t$
of all charged fragments detected in each collision:
\begin{equation}
  \vec{Q}^{\nu}~=~\sum_{i}
  \nu_{i} \left(
  \frac{\vec{V}^{t}}{|\vec{V}^{t}|} \right)_{i}
    \label{eq:qnu}
\end{equation}
where the weight $\nu_i$ depends on the pulse height of the ith charged 
particle. The values of the $\nu_i$ are positive for $\alpha \geq \alpha_0$ 
and zero for $\alpha < \alpha_0$,  where the quantity 
$\alpha \equiv (Y/Y_{p})_{CM}$ is the rapidity
$Y$ normalized to the projectile rapidity $Y_{p}$ in the center-of-mass system,
and $\alpha_{0}$ is a threshold rapidity. The rapidity $Y$ of a charged 
particle with an identified Z was calculated with an assumption that its mass
is the mass of proton, $^4He$, and $^7Li$, respectively, for     
$Z = 1,\ 2$, and $Z > 2$.  The dispersion angle 
$\Delta\phi_R$, the angle between the {\em estimated} and {\em true} 
reaction planes, is defined \cite{zha90} as 
\begin{equation}
  \cos \Delta \phi_{R}~=~\frac{\langle V_{x}'\rangle}{\langle V_{x}\rangle}
  \label{eq:cos}
\end{equation}
\noindent
where $\langle V_{x}'\rangle$ and $\langle V_{x}\rangle$ are projections of 
unit vectors onto the {\em estimated} and the {\em true} reaction plane, 
respectively.  All the results presented in this report are corrected
for this dispersion.

The smaller the dispersion angle, the closer the {\em estimated} reaction 
plane to the {\em true} one.  
By averaging the projection of the unit 
vector $\left( \frac{\vec{V}^{t}}{|\vec{V}^{t}|} \right)_{i}$ onto the {\em 
true} reaction plane over the total events, one can obtain
\begin{equation}
  \langle V_{x}\rangle~=~\left[ \frac{\overline{Q^{2}-W}}
{\overline{W(W-1)}} 
\right]^{\frac{1}{2}};~~~~\vec{Q}~=~\sum_{i} 
\omega_{i} \left( 
\frac{\vec{V}^{t}}{|\vec{V}^{t}|} \right)_{i},~~~~W = \sum_{i} \omega_{i}, 
  \label{eq:vx}
\end{equation}
\noindent
where $i$ is the particle index.  
The weight $\omega_{i}$ in the above 
equation is equal to unity for $\alpha \geq \alpha_{0}$ and is equal to zero 
for $\alpha < \alpha_{0}$. 

The average of the normalized in-plane vector of all charged particles 
projected onto the {\em estimated} reaction plane, $\langle V_{x}'\rangle$, 
can be estimated by 
\begin{equation}
  \langle V_{x}'\rangle~=~\overline{\left( \frac{\vec{V}^{t}}{|\vec{V}^{t}|} 
\right)_{i} \cdot 
\frac{\vec{Q}^{\nu}_{i}}{|\vec{Q}^{\nu}_{i}|}};~~~~\vec{Q^{\nu}_{i}}~=~\sum_{j 
\neq i} \nu_{j} \left( 
\frac{\vec{V}^{t}}{|\vec{V}^{t}|} \right)_{j}
  \label{eq:vx1}
\end{equation}
\noindent
where the weight $\nu_{j}$ depends on the pulse-height of the 
charged-particles.  
The weighting factors $\nu_j$ were chosen to minimize the dispersion angle
$\Delta\phi_R$. For different values of $\nu_j$, the dispersion angles for
La-La 
collisions at 250 and 400A MeV and Nb-Nb collisions at 400A MeV exhibit a
broad minimum (around $\alpha_0 = 0.2$) 
as a function of the threshold rapidity $\alpha_{0}$; to
minimize the dispersion, we chose $\alpha_{0} = 0.2$.

Figure~\ref{fig:wadc} shows a typical pulse-height spectrum for one of the 184 
TOF wall detectors for La-La collisions at 400A MeV.  The peak labelled $Z=1$ 
is from hydrogen isotopes;  that labelled $Z=2$ is from helium isotopes.  
The third and the following peaks labelled $Z>2$ represent charged 
particles with $Z>2$.  Based on these peaks, three different sets of 
weights were tuned to estimate the reaction plane closest to the true reaction 
plane.  Figure~\ref{fig:wt} shows the dispersion angles obtained at 
different sets of weights for La-La collisions at 400A MeV.  The uncertainties 
in this figure were not calculated because our main interest was to minimize 
the values of the dispersion angles.  In this work, ten sets of weights (as 
shown in Fig.~\ref{fig:wt}) were used to calculate the dispersion angles.  
The weights 1, 2, and 2.5 for $Z=1, 
Z=2,$ and $ Z>2$ isotopes, respectively, yielded a minimum value for the 
dispersion $\Delta \phi_{R}$ in $\phi_{R}$.  With these weights, the 
dispersion angles from equation~(\ref{eq:cos}) were found to be $31.3^{\circ} 
\pm 2.4^{\circ}$ and $33.8^{\circ} \pm 2.3^{\circ}$ for La-La collisions at 
400 and 250A MeV, respectively, and $39.8^{\circ} \pm 3.0^{\circ}$ for 
Nb-Nb collisions at 400A MeV.  Without the charged-particle identification 
($i.e.$,~~$\nu_{j}=1$), equation~(\ref{eq:vx1}) corresponds to that in 
references~\cite{au93,nb93};  the dispersion angles were about 3$^{\circ}$ 
larger in these three sets of data.  After the azimuthal angle $\phi_{R}$ of 
the reaction plane was determined, the neutron azimuthal angle relative to the 
reaction plane ($\phi - \phi_{R}$) was obtained for each event.  

\subsection{Determination of the Flow Axis} 

To study the emission patterns and the event shapes of the fragments, the 
sphericity method \cite{gfs82} was used.  
From a set of charged particles in the center-of-mass system for each event, 
a sphericity tensor is defined as
\begin{equation}
  F_{ij}~=~\sum_{\nu} \frac{1}{2m_{\nu}} {\frac{V^{\nu}_i V^{\nu}_j}{|V^{\nu}|^2}},
\label{eq:ften}
 \end{equation}
\noindent
where $m_{\nu}$ is the mass of the $\nu^{th}$ fragment, which can be one of the 
proton-like ($Z=1$), helium-like ($Z=2$), or ``lithium-like'' ($Z > 2$) 
particles identified by pulse-heights in the TOF wall as mentioned in the 
previous section. In the sphericity calculation, we omitted all tracks
in the backward rapidity range ($\alpha < 0$),
then we projected the tracks in the forward 
rapidity range ($\alpha > 0$) to the backward rapidity range, $\vec{V} 
\rightarrow - \vec{V}$. 
Our sphericity calculations reconstructed polar flow angles, which allowed
neutron squeeze-out to become visible.
By diagonalizing the flow tensor in the center-of-mass system, the event 
shape can be approximated as a prolate (or cigar-shaped) ellipsoid.  The angle 
between the major axis of the flow ellipsoid and the beam axis is defined as a 
flow angle $\theta_{F}$.  

Quantitatively, the flow angle \cite{gus84} was obtained as a polar angle 
corresponding to the maximum eigenvalues of the flow tensor.  
Figure~\ref{fig:flow} shows the distribution of the flow angle $\theta_{F}$ 
for La-La and Nb-Nb collisions at 400A MeV with the same impact
parameter. For Nb-Nb and 
La-La collisions at the same energy, the 
peak of the neutron flow angle distribution moves to a larger angle
as the mass of the system increases. This trend to larger flow angles with
increasing target-projectile mass was observed previously for charged particles
\cite{rit85} by the Plastic Ball Spectrometer, and predicted qualitatively 
from Vlasov-Uehling-Uhlenbeck calculations \cite{mhs85}.  

\section{Results}

\subsection{Neutron Triple-Differential Cross Sections}

The results of the triple-differential cross sections, 
$d^{3}\sigma/d\alpha\cdot d\cos \theta \cdot d(\phi - \phi_{R})$, for neutrons 
emitted at a polar angle $\theta$ with a normalized center-of-mass rapidity 
$\alpha \equiv (Y/Y_{p})_{CM}$ (in units of the projectile rapidity $Y_{p}$) 
are presented as a function of the azimuthal angle ($\phi - \phi_{R}$) with 
respect to the reaction plane.  The data were summed in four rapidity bins 
($\Delta \alpha$) for each detector:  backward rapidities ($-1.0 \leq \alpha < 
-0.2$), mid-rapidities ($-0.2 \leq \alpha < 0.2$), intermediate-forward 
rapidities ($0.2 \leq \alpha < 0.7$), and projectile-like rapidities ($0.7 
\leq \alpha < 1.2$).  The uncertainties in the triple-differential cross 
sections include both statistical and systematic uncertainties;  however, 
statistical uncertainties dominate systematic uncertainties \cite{ela94}.  

Figure~\ref{fig:sigy1} shows the triple-differential cross sections for 
neutrons emitted in the backward rapidity bin at a polar angle of 72$^{\circ}$ 
for La-La and Nb-Nb collisions at 400A MeV and La-La collisions at 250A MeV.  
The cross sections in this rapidity bin have a minimum at $(\phi - 
\phi_{R})=0^{\circ}$ and a maximum at $(\phi - \phi_{R})=\pm 180^{\circ}$.  
Similar characteristics can be found at other polar angles, ranging from 
3$^{\circ}$ to 80$^{\circ}$.  

Figure~\ref{fig:sigy2} shows the neutron triple-differential cross sections for 
the mid-rapidity bin at one of the polar angles 
({\em viz.},~~$\theta=48^{\circ}$) for La-La and Nb-Nb collisions at 400A MeV 
and La-La collisions at 250A MeV.  In this bin, the neutrons are aligned 
perpendicular to the reaction plane at $(\phi - \phi_{R})=\pm 90^{\circ}$ (so 
called {\em squeeze-out}) \cite{gut89,gut90,lei93};  however, this effect is 
barely noticeable in our figures.  Rotating the event onto the flow axis 
enhances our ability to see the squeeze-out effect \cite{gut89,har94}.

Figure~\ref{fig:sigy4} shows the neutron triple-differential cross sections for 
the projectile rapidity bin at a polar angle 16$^{\circ}$ (as an example) for 
La-La and Nb-Nb collisions at 400A MeV and La-La collisions at 250A MeV.  At 
each polar angle, the azimuthal distribution for this rapidity bin peaks at 
$(\phi - \phi_{R})=0^{\circ}$ and has a minimum at $(\phi - \phi_{R})=\pm 
180^{\circ}$.  The peak at 0$^{\circ}$ is the result of the side-splash and 
bounce-off effects, where bounce-off cannot be separated clearly from the 
side-splash.  The resulting cross sections in these projectile rapidities also 
have contributions from both participants and spectators like those at target 
or backward rapidities;  however, the side-splash dominates at the larger 
polar angles where the collisions are more central because of the multiplicity 
cut chosen for this analysis.  

Finally, Fig.~\ref{fig:sigy3} shows the neutron triple-differential cross 
sections for the intermediate-forward rapidity bin at the polar angle 
20$^{\circ}$ for La-La and Nb-Nb collisions at 400A MeV and La-La collisions at 
250A MeV.  The distributions at small polar angles in this rapidity bin 
reflect the distributions in the mid-rapidity bin.  The larger the polar 
angles, the more noticeable the peaks at $(\phi - \phi_{R})=0^{\circ}$ and the 
distributions become more like the distributions from the projectile rapidity 
regions.  

\subsection{Average In-plane Momentum} \label{sec:px}

Figure~\ref{fig:px400} shows the mean transverse momentum projected into the 
reaction plane, $\langle P_{x} \rangle$, for neutrons as a function of 
normalized center-of-mass rapidity, $\alpha = (Y/Y_{p})_{CM}$, for 
La-La collisions at an energy of 400 and 
250A MeV.  The technique of extracting the $\langle P_{x}\rangle$ for 
neutrons can be found in reference~\cite{au93}.  

The data display the typical S-shaped behavior as described by Danielewicz and 
Odyniec \cite{do85}.  
Neutrons in the low-energy regions (below $\approx$ 55 MeV) 
were not included in order to eliminate background contamination;  thus, the 
curve is not completely symmetric and 
the slope of the average in-plane transverse momentum at negative rapidities
is steeper than that at positive rapidities because the cut on the low energy
neutrons rejects neutrons with low transverse momenta at negative rapidities.
From 
Fig.~\ref{fig:px400}, the average in-plane momentum increases with increasing 
bombarding energies. 

In the positive rapidity region, the $\langle P_x \rangle$ vs 
$\alpha$ curves are 
straight lines up to $\alpha = 0.5$. We extracted the slope at mid-rapidity
(up to $\alpha = 0.5$) with a linear fit to $\langle P_x \rangle$ in the 
region 
unaffected by the cut on the neutron energy; Doss {\it et al.}\cite{dos86}
defined this slope as the {\it flow} F. 
Because the flow is 
determined at mid-rapidity, it has contributions only from the participants.  
The flows found in this analysis are $145 \pm 11$ and $104 \pm 13$ MeV/c from
La-La collisions at 400 and 250A MeV, respectively.
From these data, we see that the neutron 
flow increases with increasing beam energy. Previously, 
Doss {\it et al.}\cite{dos86} observed that the flow of charged particles
increases with beam energy, reaches a broad maximum at about 650A MeV, and
falls off slightly at higher energies.

\subsection{Neutron Squeeze Out} \label{sec:sqz}

One of the results of this experiment is the observation of neutrons emitted
preferentially
in a direction perpendicular to the reaction plane. In their paper reporting
this component of collective flow for charge particles,
Gutbrod {\it et al.}\cite{gut89} called this collective phenomenon {\it 
out-of-plane squeeze out.}

To see neutron squeeze out, we performed two coordinate rotations:
First, we rotated the x coordinate around the beam or z-axis to align it
with the summed transverse-velocity vector $\vec{Q}$ given by 
equation~(\ref{eq:vx}); second, we rotated the z-axis around the y-axis
by the flow angle $\Theta_F$. After this second rotation, the $z'$-axis
of the new $x'y'z'$ coordinates, where $y'=y$, is on the major axis of
flow ellipsoid. Then, for neutrons with transverse momenta
$p'_{z}$ in the region $-0.1 \leq p'_{z} < 0.1$, where 
$p'_{z}=(P'_{z}/P'_{proj})_{CM}$, neutron squeeze out became visible as
a peak at azimuthal angles $\phi'~=~\pm 90^{\circ}$ for the distribution
in the $x'y'$-plane. 
Figure~\ref{fig:squez} shows the neutron azimuthal distributions in the 
$x'y'$-coordinates or neutron squeeze-out of three systems: La-La and Nb-Nb 
collisions at 400A MeV and La-La collisions at 250A MeV. 
The spectator neutrons are excluded from Fig.~\ref{fig:squez} by
removing neutrons of high and low momenta at small angles.
By excluding spectator neutrons \cite{dos85} emitted from the
projectile and the target, neutrons evaporated \cite{mad85,mad90} from an
excited projectile at small polar angles ($\theta \leq 8^{\circ}$) were
excluded also. 
Squeeze out of neutrons was observed previously\cite{lei93} for Au-Au
collisions at 400A MeV.

\section{Theoretical Comparisons}

Welke {\em et al.} \cite{wel88} examined the sensitivity of $d\sigma/d\phi$ to 
the nuclear matter equation of state.  Because there was a net flow in the 
projectile and target rapidity bins, Welke {\em et al.} described the shapes 
of the azimuthal distribution by the ratio of the maximum cross section 
$(d\sigma/d\phi)_{max}$ to the minimum cross section $(d\sigma/d\phi)_{min}$.  

For each polar angle, the cross sections measured in the experiment are fitted 
with the function, $\sigma_{3}(\phi - \phi_{R}, 
\theta)~=~a(\theta)~\pm~b'(\theta)~\cos (\phi - \phi_{R})$, 
where the $+(-)$ sign stands for positive (negative) rapidity bin, and 
$b'(\theta)=b(\theta) e^{-(\Delta \phi_{R})^{2}/2}$ is the correction 
for a finite rms dispersion $\Delta \phi_{R}$.  
For positive rapidity 
particles, the cross sections peak at $(\phi - \phi_{R})=0^{\circ}$ and 
deplete at $(\phi - \phi_{R})=\pm 180^{\circ}$, as seen in 
Fig.~\ref{fig:sigy4}.  The maximum azimuthal anisotropy for positive 
rapidity neutrons becomes $ r(\theta)~=~\left[ a(\theta)~+~b(\theta)\right]/
\left[ a(\theta)~-~b(\theta)\right]$.  Figure~\ref{fig:smr} is the 
polar-angle-dependent maximum azimuthal anisotropy ratio $r(\theta)$ for the 
four sets of data as indicated;  the data for Au-Au collisions at 400A MeV are 
from Elaasar {\em et al}. \cite{ela94}.  From Fig.~\ref{fig:smr}, the maximum 
azimuthal anisotropy appears to be independent of both the mass of the 
colliding nuclei and the beam energy.  

For theoretical comparison, we used the Boltzmann-Uehling-Uhlenbeck (BUU) 
approach \cite{bd88} with a parameterization of a momentum-dependent nuclear 
mean field suggested in reference \cite{wel88}.  In the calculations, the 
incompressibility modulus $K$ was set to be 215 MeV, and the contributions to 
the cross sections from composite fragments were subtracted by rejecting 
neutrons when the distance between the neutron and another nucleon from the 
same BUU ensemble \cite{bd88} was less than a critical 
value $d_c$ \cite{ab85}.  
Within a given BUU run, a nucleon was considered {\em free} only if no other 
nucleons were found within the critical distance $d_c$.  We are 
justified in restricting our coalescence criterion to coordinate 
space as nucleons that are far apart in momentum space will have drifted
away from each other. Our analysis is performed at the time that 
the momentum space distributions begin to freeze-out\cite{zg94}.
Also we have verified quantitatively the soundness of this
argument by performing a full six-dimensional coalescence.
It is well known from transport theory calculations that the 
transverse momentum generation in heavy-ion reactions begins quite early in 
the history of the reaction and then saturates \cite{gal90}.  For {\em free} 
neutrons, the critical distance $d_c$ was chosen to be 2.8~fm for both Nb-Nb 
and La-La at 400A MeV and 3.0 fm for La-La at 250A MeV by fitting the 
polar-angle dependence of the double-differential cross sections.  The 
double-differential cross sections in the rapidity $0.7 \leq \alpha < 1.2$ bin 
for La-La and Nb-Nb collisions at beam energy 400A MeV and La-La collisions at 
250A MeV are shown in Fig.~\ref{fig:dcr_buu}.  The filled symbols represent 
the data, and the open symbols represent the BUU calculations.  The BUU 
calculations (with $d_c = 0$~fm) of the double differential cross sections are 
significantly higher than the data because the data do not include the 
composite fragments;  in other words, the data contain {\em free} neutrons 
only.  For $d_c = 2.8$~fm, the BUU prediction is lower than the data at 
small polar angles because the BUU calculations do not treat neutron 
evaporation which occurs at polar angles below $\sim 9^{\circ}$, as observed 
previously by Madey {\em et al}. \cite{mad90}.  

By restricting the BUU calculations to {\em free} neutrons with $K = 215$~MeV, 
the triple-differential cross sections are compared with data.  
Figures~\ref{fig:sigy1}~--~\ref{fig:sigy3} show triple-differential cross 
sections for four rapidity regions:   $-1.0 \leq \alpha < -0.2$, $-0.2 \leq 
\alpha < 0.2$, $0.7 \leq \alpha < 1.2$, $0.2 \leq \alpha < 0.7$.  In these 
figures, the open symbols represent the BUU calculations.  The solid line in 
each figure represents the data corrected to zero dispersion $\Delta \phi_{R} 
= 0$. 
In these calculations, the negative $(\phi - 
\phi_{R})$ region is symmetric with respect to the positive side.  In 
Fig.~\ref{fig:sigy1}, the BUU results are higher than the data beyond 
$(\phi - \phi_{R}) = 90^{\circ}$.  In Fig.~\ref{fig:sigy2}, the BUU results 
tend to peak at $(\phi - \phi_{R}) = 90^{\circ}$;  in other words, the BUU 
results show the characteristics of the out-of-plane squeeze-out effect for 
{\em free} neutrons in the mid-rapidity bin, but it is very hard to see the 
squeeze-out phenomena in the rapidity coordinates (see 
section~\ref{sec:sqz}).  In Figs.~\ref{fig:sigy4} and \ref{fig:sigy3}, the BUU 
results generally agree with the data in these positive rapidity bins.  

The BUU calculations of the polar-angle dependence of the maximum azimuthal 
anisotropy ratio $r(\theta)$ for {\em free} neutrons emitted from Nb-Nb, La-La, 
and Au-Au collisions at 400A MeV and La-La collisions at 250A MeV are shown in 
Fig.~\ref{fig:r400buu} for $(b_{max}/2R) = 0.5$.  The BUU results for Au-Au 
collisions at 400A MeV are from Elaasar {\em et al}. \cite{ela94}.  In this 
figure, the BUU calculations with $K = 380,~215,~150$ MeV were carried out for 
{\em free} neutrons (with $d_c = 2.8$ fm for Nb-Nb and La-La collisions and $d_c = 
3.2$ fm for Au-Au collisions).  Consistent with the experimental data (see 
Fig.~\ref{fig:smr}), the BUU calculations in Fig.~\ref{fig:r400buu} show 
little sensitivity to the mass and the beam energy.  The BUU calculations of 
the polar-angle-dependent maximum azimuthal anisotropy ratio $r(\theta)$ for 
{\em free} neutrons emitted from La-La and Nb-Nb collisions at 400A MeV and 
La-La collisions at 250A MeV are compared with the data in 
Fig.~\ref{fig:smr_buu}.  The multiplicity cuts indicated in this figure 
correspond to the ratio of the maximum impact parameter to the nuclear radius 
$(b_{max}/2R) = 0.5$.  The filled and open symbols in this figure represent 
the data and the BUU calculations (with $K = 150,~215,~380$ MeV), 
respectively.  Both the experimental results and the BUU calculations were for 
zero dispersion $\Delta \phi_{R} = 0$.  As one can see from this figure, the 
polar-angle dependence of the maximum azimuthal anisotropy ratio $r(\theta)$ 
is insensitive to the incompressibility modulus $K$ in the nuclear equation of 
state. This insensitivity was noted also in Ref\cite{zg94}.  

Figure~\ref{fig:px} is the comparison between data and the BUU calculations 
with $K = 215$ MeV for in-plane transverse momentum $\langle P_{x} \rangle$ 
for {\em free} neutrons emitted from La-La and Nb-Nb collisions at 400A MeV 
and La-La collisions at 250A MeV.  Similar to the other figures, the filled 
symbols and open symbols represent the data and the BUU calculations, 
respectively.  The BUU calculations generally agree with the data within their 
uncertainties, especially in the mid-rapidity region which gives the 
information about flow in the unit of MeV/c (see section~\ref{sec:px}).  

\begin{figure}
  \caption{Neutron squeeze-out in the mid-momentum ($-0.1 \leq p_{z}' = 
(P_{z}'/P_{proj}')_{CM} \leq 0.1$) region for La-La collisions at 400A MeV and 
250A MeV, and Nb-Nb collisions at 400A MeV.  The solid lines represent the 
data.  The dotted, dashed, and dot-dashed lines represent BUU theory with 
$K = 380,~215,$ and 150~MeV, respectively, for {\em free} neutrons with a 
critical distance $d_c=2.8$~fm.  All three lines are almost on top of each other 
in La-La collisions at 250A MeV.  
\label{fig:squez_buu}}
\end{figure}
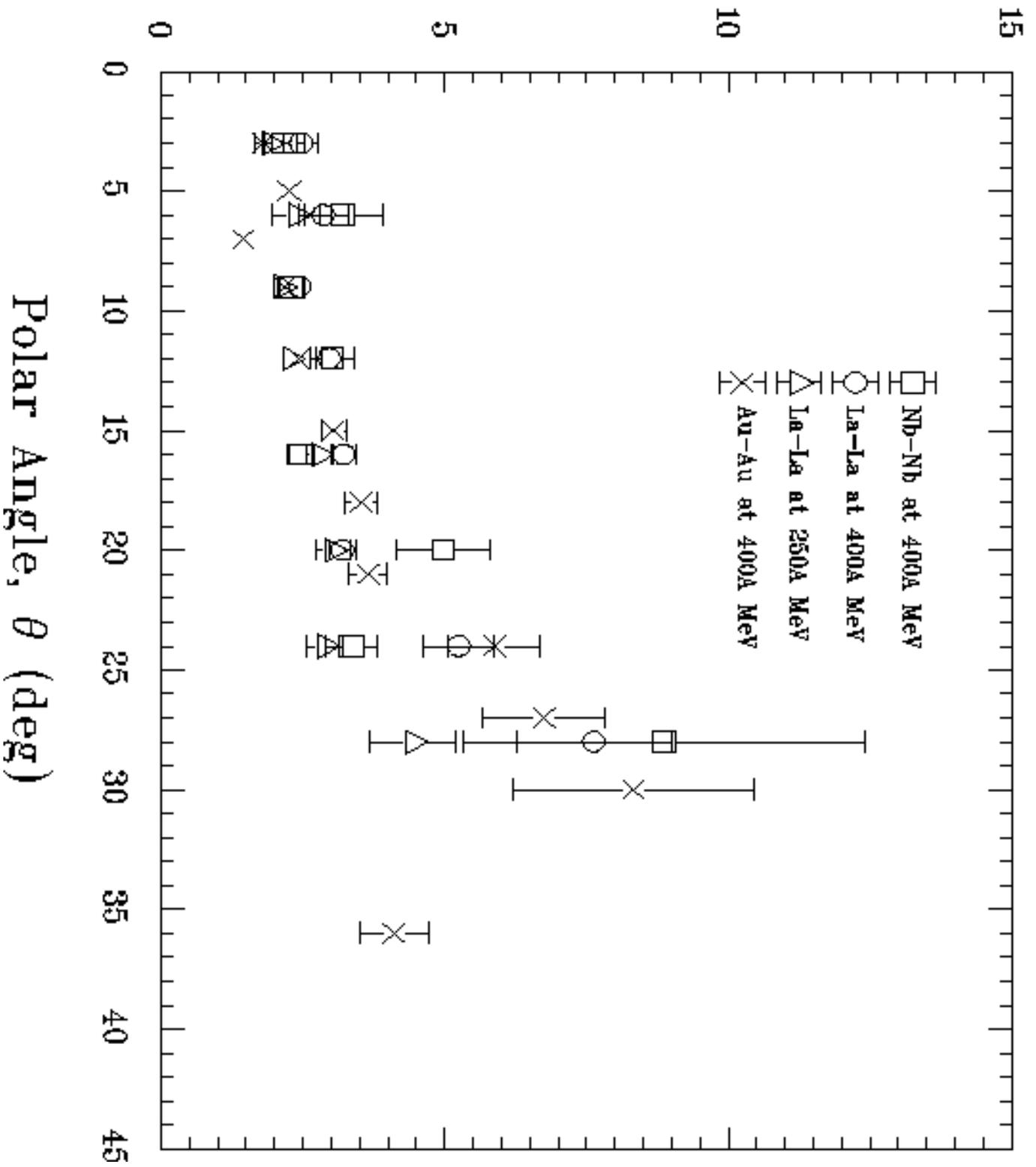
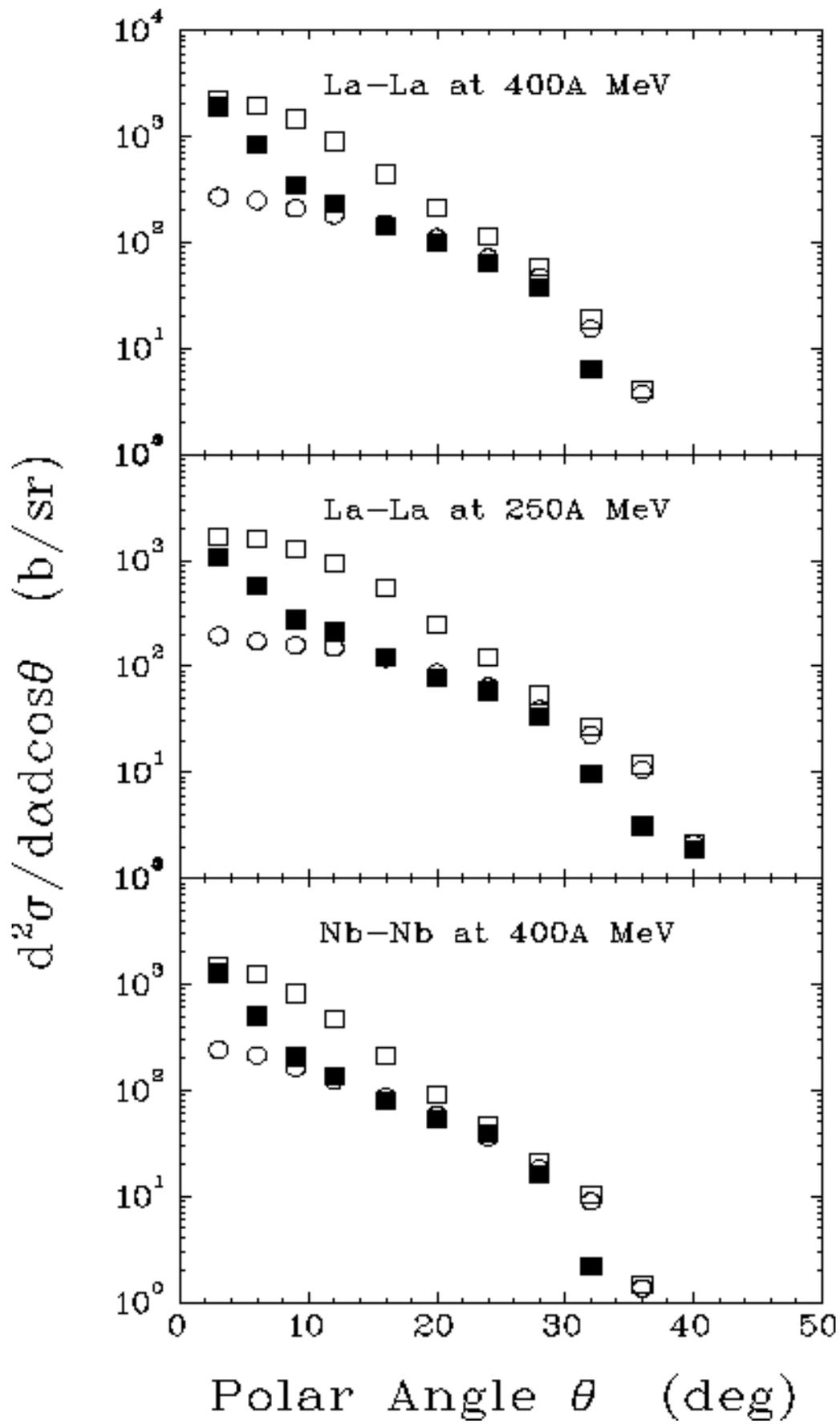
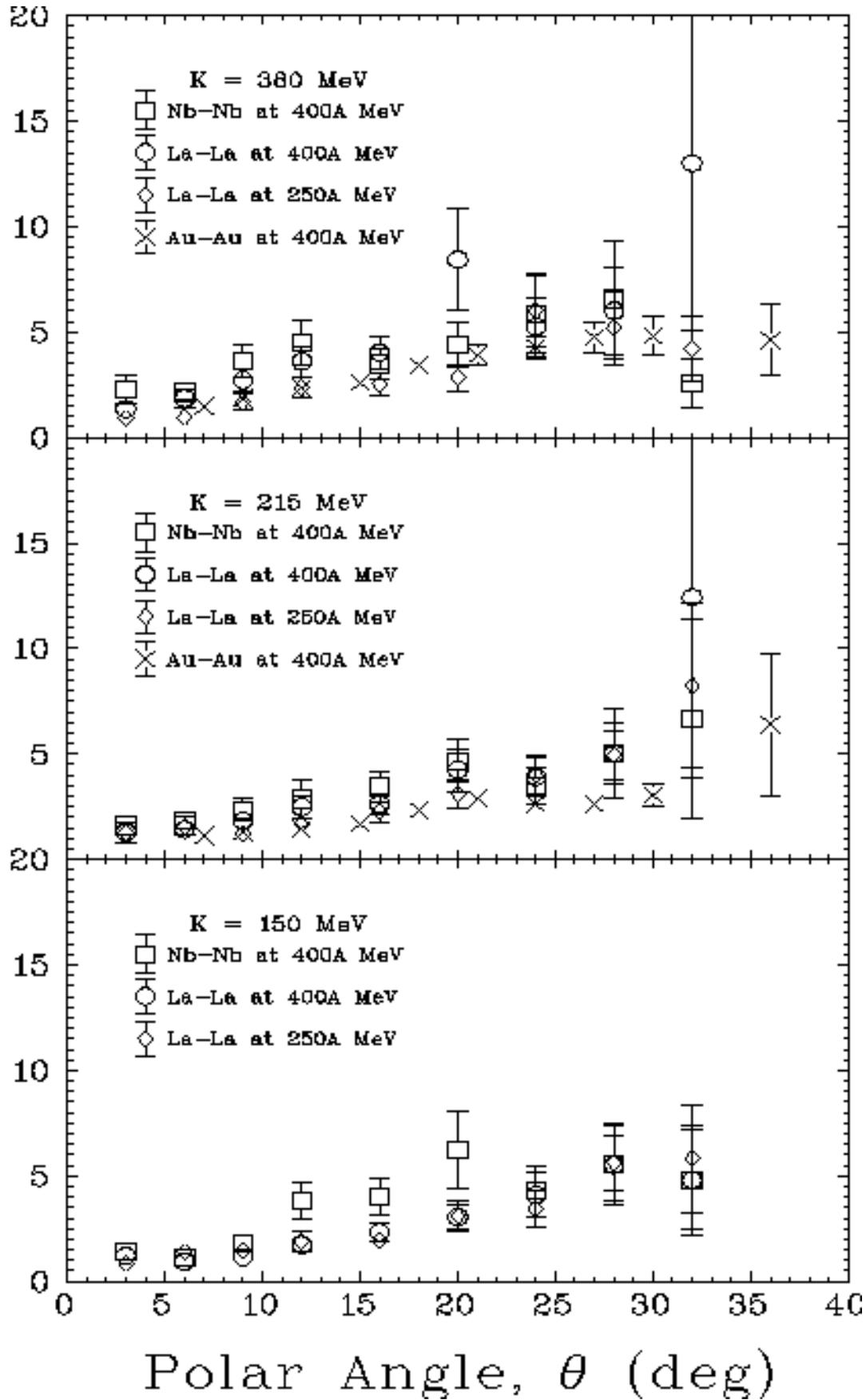
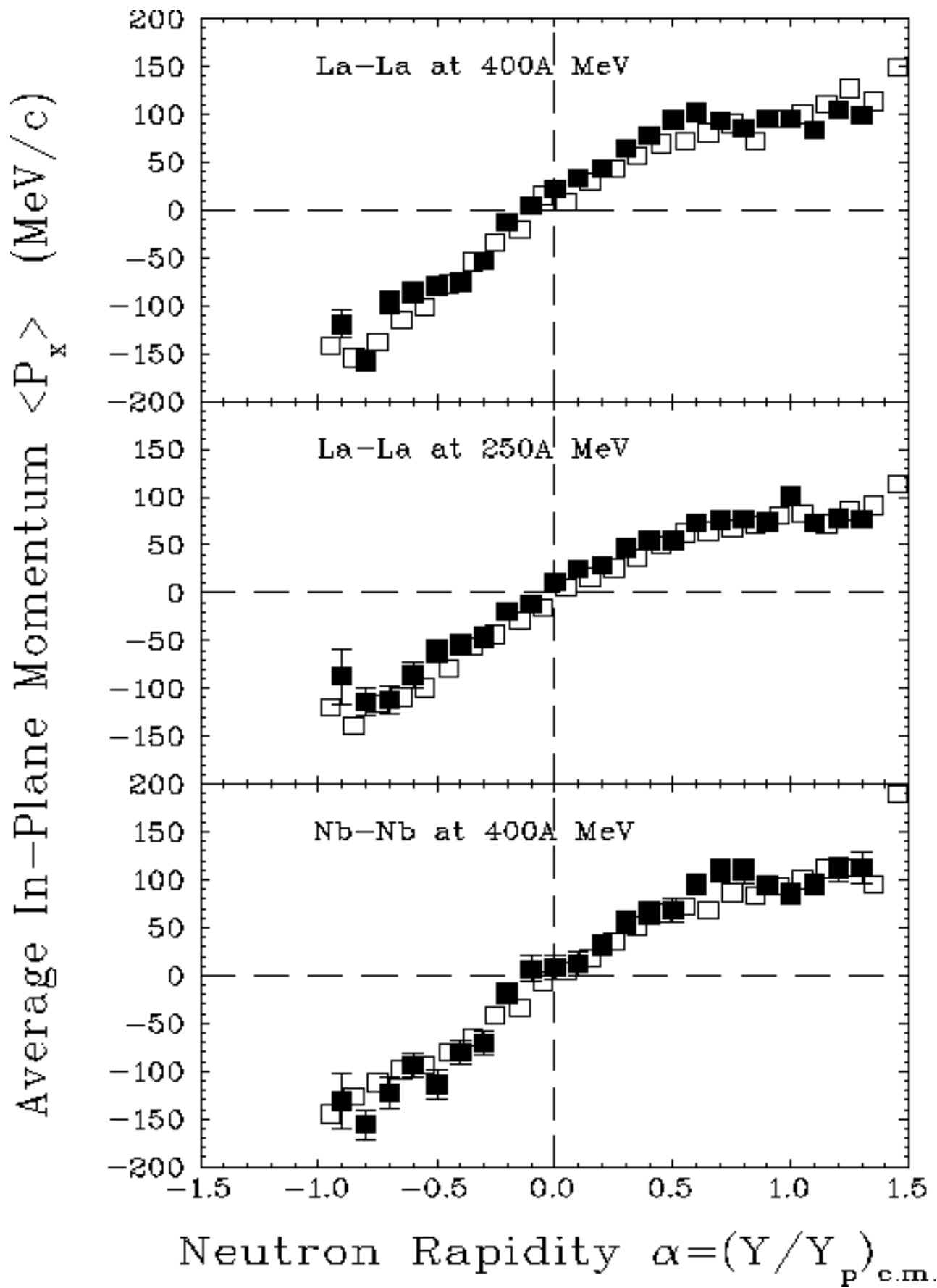
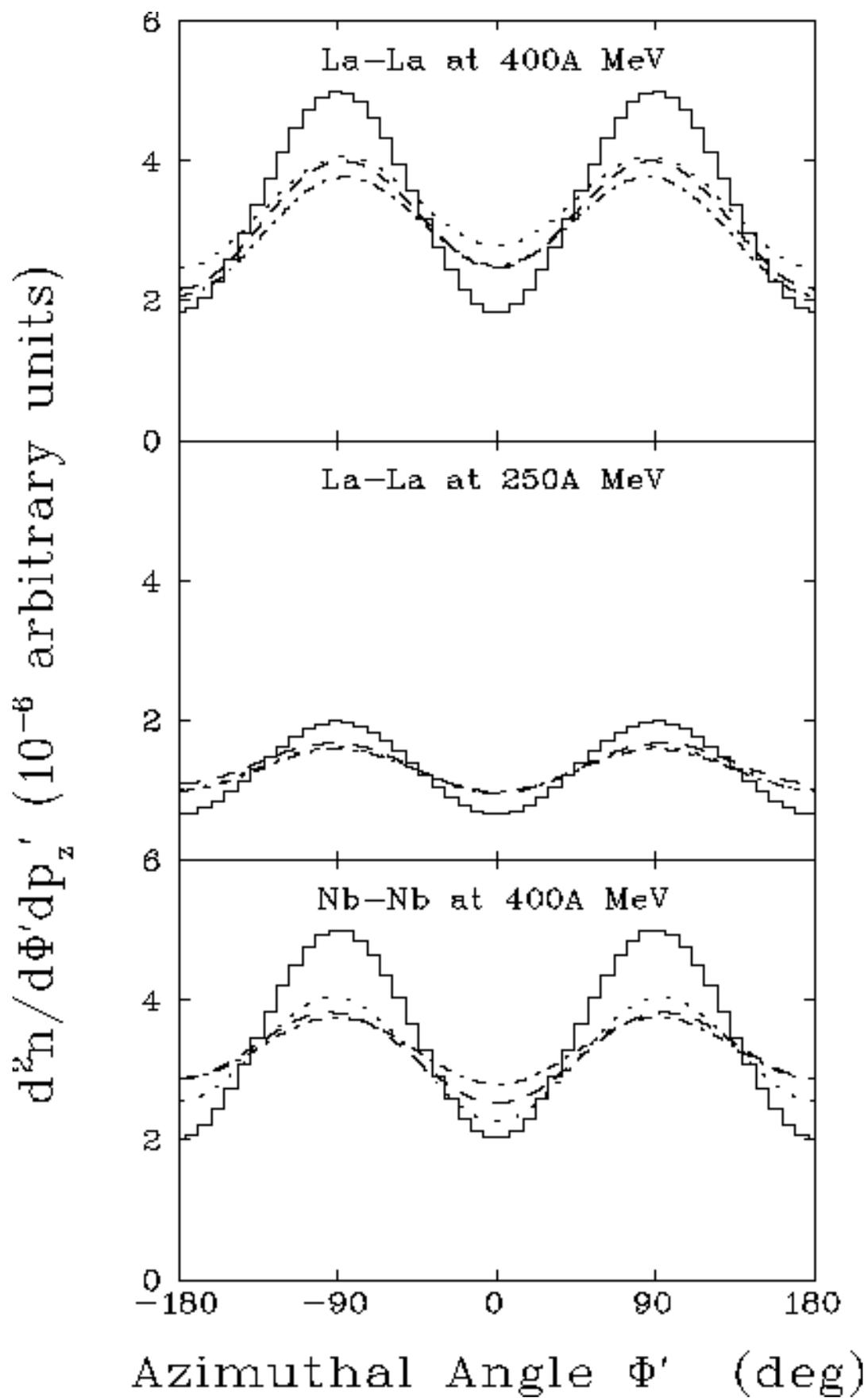

Another observable obtained from BUU calculations is the out-of-plane 
squeeze-out of {\em free} neutrons.  The comparisons are depicted in 
Fig.~\ref{fig:squez_buu} for La-La and Nb-Nb collisions at 400A MeV and La-La 
collisions at 250A MeV. The solid lines represent the data.  The dotted, 
dashed, and dot-dashed lines represent the BUU calculations with 
$K = 380,~215,$ and 150~MeV, respectively.  All three lines are almost on top 
of each other in La-La collisions at 250A MeV.  The squeeze out of
neutrons in the 
normalized momentum coordinates in the center-of-mass system is compared with 
the BUU model.  Both the experimental results and the BUU calculations are 
for zero dispersion $\Delta \phi_{R} = 0^{\circ}$.
After correcting to zero dispersion in the 
experimental results, the neutron squeeze-out becomes larger.  
It can be seen from 
Fig.~\ref{fig:squez_buu} that the squeeze-out of {\em free} neutrons 
from BUU calculations is insensitive to the incompressibility modulus $K$.  
From the comparison between the BUU calculations and the neutron squeeze-out 
in Fig.~\ref{fig:squez_buu}, the squeeze-out effect from the experimental data 
in this work is significantly stronger than that from the BUU calculations with 
$K = 380,~215,~150$ MeV.  It remains to be seen whether the apparent 
disagreement between the BUU squeeze-out results and the data persist as 
greater statistics and accuracy are reached for the calculations. We estimate 
the present uncertainties to be of the order of 25\%. 
Statistically meaningful calculations of squeeze-out in BUU remain challenging
as the effect truly needs to be established at a level past fluctuations.

\section{Conclusions}
We measured triple-differential cross sections of neutrons emitted at
several polar angles in multiplicity-selected La-La collisions at 
250 and 400A MeV (and Nb-Nb collisions at 400A MeV) as a function of 
the azimuthal angle with respect to the reaction plane. We compared the
measured cross sections with BUU calculations for free neutrons; the
BUU calculations (with an incompressibility modulus $K = 215$ MeV) 
agree with the measured cross sections, except for the smallest polar
angle where the BUU calculations do not treat neutron evaporation. 

The La-La data at 400A MeV permitted us to extend our ealier studies of
the maximum azimuthal anisotropy ratio as a probe of the nuclear equation
of state, and to conclude that the maximum azimuthal anisotropy ratio 
is insensitive to the beam energy from 250 to 400A MeV for the La-La system. 
The uncertainties in the measurement of the maximum azimuthal anisotropy 
ratio are about 15\% which do not allow us to investigate its dependence on the 
mass of the projectile-target system.  BUU calculations
confirm the lack of sensitivity of the maximum azimuthal anisotropy ratio
to the mass of the colliding system and to the beam energy. BUU calculations
suggest also that the maximum azimuthal anisotropy ratio is insensitive to
the incompressibility modulus K in the nuclear equation of state.

The flow of neutrons was extracted from the slope at mid-rapidity of the
curve of the average in-plane momentum verses the neutron center-of-mass
rapidity. The flow of neutrons emitted in La-La collisions at 250 and 400A
MeV increases with beam energy. BUU calculations with with an 
incompressibility modulus $K = 215 $ MeV for free neutrons agree 
generally with the data.

We observed the preferential emission of neutrons in a direction perpendicular
to the reaction plane in three systems [viz. 400 and 250A MeV La-La, and
400A MeV Nb-Nb]. This component of collective flow, observed first for charged
particles is known as out-of-plane squeeze out. BUU calculations of 
out-of-plane squeeze out of free neutrons are insensitive to the
incompressibility modulus K for values of 150, 215, and 380 MeV. After 
correcting the experimental results to zero dispersion, the observed 
squeeze out of neutrons is significantly stronger than that from BUU theory.

\acknowledgements

This work was supported in part by the National Science Foundation under Grant 
Nos.\ PHY-91-07064, PHY-88-02392, and PHY-86-11210, the U.S. Department of 
Energy under Grant No.\ DE-FG89ER40531 and DE-AC03-76SF00098, and the 
Natural Sciences and Engineering Research Council of Canada and by the Fonds 
FCAR of the Quebec Government. \\


\newpage
\begin{table}[p]

\caption{Neutron detector placement and shadow shield configurations.  
\label{tab:neuss}}

\begin{tabular}{lccccccccc}
\multicolumn{2}{c}{Detector} & Polar & Flight & Distance from &
\multicolumn{4}{c}{Configurations} \\
No. & Width & Angle & Path & Target to & A & B & C & D & E \\
  & w (in)& $\theta$ (deg)& x (m) & Front of Shield (m) & &  &  &  &  \\ \hline
N16 & 20 &  80.0 & 6.00 & 3.0 &  & $SS4$ &     &     &     \\
N14 & 20 &  64.0 & 7.00 & 3.5 &  &     & $SS4$ &     &     \\
N12 & 20 &  48.0 & 8.00 & 4.5 &  &     &     & $SS4$ &     \\
N10 & 10 &  36.0 & 8.40 & 4.8 &  &     &     &     & $SS3$ \\
N8  & 10 &  28.0 & 8.40 & 4.8 &  & $SS3$ &     &     &     \\
N6  & 10 &  20.0 & 8.40 & 5.0 &  &     & $SS3$ &     &     \\
N4  &  5 &  12.0 & 8.40 & 5.0 &  &     &     & $SS2$ &     \\
N2  &  1 &   5.9 & 8.39 & 5.0 &  &     &     &     & $SS1$ \\
N1  &  1 &  -3.0 & 8.40 & 5.0 &  & $SS1$&  &     &     \\
N3  &  5 &  -9.0 & 8.40 & 5.0 &  &     & $SS2$&  &     \\
N5  & 10 & -16.0 & 8.40 & 5.0 &  &     &     &$SS3$&   \\
N7  & 10 & -24.0 & 8.40 & 5.0 &  &     &     &     & $SS3$ \\
N9  & 10 & -32.0 & 8.40 & 4.8 &  & $SS3$ &     &     &     \\
N11 & 20 & -40.0 & 8.40 & 4.8 &  &     & $SS4$&  &     \\
N13 & 20 & -56.0 & 7.50 & 3.8 &  &     &     & $SS4$ &     \\
N15 & 20 & -72.0 & 6.20 & 3.1 &  &     &     &     & $SS4$ \\ [3mm]
Type &\multicolumn{2}{c}{Dimensions ($in^{3}$)}&Number&  &  &  &  &  &  \\
$SS1$ &\multicolumn{2}{c}{$5.5\times 40\times 40$}&1& &0&1&0&0&1 \\
$SS2$ &\multicolumn{2}{c}{$8.0\times 36\times 43$}&1& &0&0&1&1&0 \\
$SS3$ &\multicolumn{2}{c}{$14\times 40 \times 42$}&2& &0&2&1&1&2 \\
$SS4$ &\multicolumn{2}{c}{$20 \times 40\times 42$}&2& &0&1&2&2&1 \\ 
\end{tabular}

\end{table}





\newpage
\begin{figure}
  \caption{Experimental arrangement.  
\label{fig:expt}}
\end{figure}

\begin{figure}
  \caption{Charged-particle multiplicities with and without the target in place 
for La-La collisions at 400 and 250A MeV, and Nb-Nb collisions at 400A MeV.  
\label{fig:mult}}
\end{figure}

\begin{figure}
  \caption{Pulse-height spectrum from 400A MeV La-La collisions for one of the 
184 detectors in the time-of-flight wall. 
\label{fig:wadc}}
\end{figure}

\begin{figure}
  \caption{Dispersion $\Delta \phi_{R}$ in the azimuthal reaction plane angle 
$\phi_{R}$ with different sets of weights tuned for $Z=1$, $Z=2$, and $Z>2$ 
isotopes for La-La collisions at 400A MeV.  Each set of weights consists of 
three numbers assigned, respectively, to the charged-particles with $Z=1$, 
$Z=2$, and $Z>2$.  Among ten sets of weights used to calculate dispersion 
angle $\Delta \Phi_{R}$, the set (1, 2, 2.5) gives a minimum dispersion angle.  
\label{fig:wt}}
\end{figure}

\begin{figure}
  \caption{Distribution of the flow angle ($\theta_{F}$) for La-La 
and Nb-Nb collisions at 400A MeV with the same impact 
parameter (in the units of the nuclear radius) $(b/2R) = 0.51$. 
\label{fig:flow}}
\end{figure}

\begin{figure}
  \caption{Triple-differential cross sections for the emissions of neutrons in 
the backward rapidity bin ($-1.0 \leq \alpha < -0.2$) at a polar angle of 
$\theta = 72^{\circ}$ from La-La collisions at 400 and 250A MeV, and Nb-Nb 
collisions at 400A MeV.  The open symbols represent BUU theory with 
$K = 215$~MeV for {\em free} neutrons with a critical distance $d_c=2.8$~fm.  
The filled symbols represent the data, and the solid line represents the data 
corrected to zero dispersion $\Delta \phi_{R} = 0$.  
\label{fig:sigy1}}
\end{figure}

\begin{figure}
  \caption{Triple-differential cross sections for the emission of neutrons in 
the mid-rapidity bin ($-0.2 \leq \alpha < 0.2$) at a polar angle of 
48$^{\circ}$ from La-La collisions at 400A MeV and 250A MeV, and Nb-Nb 
collisions at 400A MeV.  The open symbols represent BUU theory with 
$K = 215$~MeV for {\em free} neutrons with a critical distance $d_c=2.8$~fm.  
The filled symbols represent the data and the solid line represents the data 
corrected to zero dispersion $\Delta \phi_{R} = 0$.  
\label{fig:sigy2}}
\end{figure}

\begin{figure}
  \caption{Triple-differential cross sections for the emission of neutrons in 
the projectile rapidity bin ($0.7 \leq \alpha < 1.2$) at a polar angle of 
16$^{\circ}$ from La-La collisions at 400A MeV and 250A MeV, and Nb-Nb 
collisions at 400A MeV.  The open symbols represent BUU theory with 
$K = 215$~MeV for {\em free} neutrons with a critical distance $d_c=2.8$~fm.  
The filled symbols represent the data and the solid line represents the data 
corrected to zero dispersion $\Delta \phi_{R} = 0$.  
\label{fig:sigy4}}
\end{figure}

\begin{figure}
  \caption{Triple-differential cross sections for the emission of neutrons in 
the intermediate-forward rapidity bin ($0.2 \leq \alpha < 0.7$) at a polar 
angle of 20$^{\circ}$ from La-La collisions at 400A MeV and 250A MeV, and Nb-Nb 
collisions at 400A MeV.  The open symbols represent BUU theory with 
$K = 215$~MeV for {\em free} neutrons with a critical distance $d_c=2.8$~fm.  
The filled symbols represent the data and the solid line represents the data 
corrected to zero dispersion $\Delta \phi_{R} = 0$.  
\label{fig:sigy3}}
\end{figure}

\begin{figure}
  \caption{Average in-plane momentum projected into the reaction plane for 
neutrons from La-La at 400 and
250A MeV as a function of the normalized rapidity of the neutrons in the 
center-of-mass system.  
\label{fig:px400}}
\end{figure}

\begin{figure}
  \caption{Azimuthal distribution ($\phi'$) of the neutrons around the flow 
axis or neutron squeeze-out in the momentum region ($-0.1 \leq p_{z}' = 
(P_{z}'/P_{proj}')_{CM} \leq 0.1$) for La-La collisions at 400 and 
250A MeV, and Nb-Nb collisions at 400A MeV.    
\label{fig:squez}}
\end{figure}

\begin{figure}
  \caption{The polar-angle-dependent maximum azimuthal anisotropy ratio 
$r(\theta)$ as a function of the polar angle $\theta$ for La-La and Nb-Nb 
collisions at 400A MeV and La-La collisions at 250A MeV in the projectile 
rapidity bin ($0.7 \leq \alpha < 1.2$).  The data for Au-Au collisions at 400A 
MeV is reproduced from Elaasar et al \protect\cite{ela94}.  
\label{fig:smr}}
\end{figure}

\begin{figure}
  \caption{Polar-angle dependence of the double-differential cross sections for 
neutrons emitted with rapidities $0.7 \leq \alpha < 1.2$ from La-La collisions 
at 400 and 250A MeV, and Nb-Nb collisions at 400A MeV.  The filled squares 
represent the data and the open symbols represent BUU theory with $K = 
215$~MeV:  open squares represent {\em all} neutrons and open circles represent 
neutrons that are not in clusters or {\em free} neutrons.  
\label{fig:dcr_buu}}
\end{figure}

\begin{figure}
  \caption{The polar-angle-dependent maximum azimuthal anisotropy ratio 
$r(\theta)$ as a function of the polar angle $\theta$ for La-La and Nb-Nb 
collisions at 400A MeV and La-La collisions at 250A MeV from BUU calculations 
with $K = 380,~215,~150$ MeV.  The value $d_c=2.8$~fm was used for these 
systems.  The results for Au-Au collisions at 400A MeV with $d_c=3.2$~fm is 
reproduced from Elaasar et al \protect\cite{nb93}.  
\label{fig:r400buu}}
\end{figure}

\begin{figure}
  \caption{The polar-angle-dependent maximum azimuthal anisotropy ratio 
$r(\theta)$ as a function of the polar angle $\theta$ for La-La collisions at 
400 and 250A MeV, and Nb-Nb collisions at 400A MeV.  The filled squares 
represent the data and the open symbols represent BUU theory with 
$K = 380$~MeV, $K = 215$~MeV, and $K = 150$~MeV as indicated in the figure.  
\label{fig:smr_buu}}
\end{figure}

\begin{figure}
  \caption{Average in-plane momentum projected into the reaction plane for 
neutrons from La-La collisions at 400A MeV and 250A MeV, and Nb-Nb collisions 
at 400A MeV as a function of the normalized rapidity of the neutrons in the 
center-of-mass system.  The filled squares represent the data and the open 
symbols represent BUU theory with $K = 215$~MeV for {\em free} neutrons with a 
critical distance $d_c=2.8$~fm.  
\label{fig:px}}
\end{figure}

\end{document}